\documentclass{ws-rv9x6}
\usepackage{subfigure}   
\usepackage{ws-rv-thm}   
\usepackage{ws-rv-van}   
\usepackage{chemmacros}
\makeindex

\newcommand{\be}{\begin{equation}}
\newcommand{\ee}{\end{equation}}

\graphicspath{{figure/}} 

\begin{document}

\chapter[Thermodynamic bounds for chemical kinetics]
{A case study of thermodynamic bounds for chemical kinetics}\label{ra_ch1}

\author{K. Proesmans}
\address{Hasselt University, B-3590 Diepenbeek, Belgium\\Coll\`{e}ge de France,	75005 Paris, France}
\author{L. Peliti}
\address{ Santa Marinella Research Institute, 00058 Santa Marinella (RM), Italy}
\author[K. Proesmans, L.Peliti and D. Lacoste]{D. Lacoste \footnote{david.lacoste@espci.fr.}}
\address{Laboratoire de Physico-Chimie Th\'eorique -- UMR CNRS Gulliver 7083,\\ 
PSL Research University, ESPCI,\\ 10 rue Vauquelin, F-75231 Paris, France}

\begin{abstract}
In this chapter, we illustrate recently obtained thermodynamic bounds for a number of 
enzymatic networks by focusing on simple examples of  
unicyclic or multi-cyclic networks. We also derive complementary relations which constrain 
the fluctuations of first-passage times to reach a threshold current.
\end{abstract}
\body


\section{Introduction}

There are generally several parameters which determine the performance of a thermodynamic system. 
One parameter is the average output flux delivered by the system, which is related to its output power. 
Another parameter is the dissipation, which may be viewed as a cost for operating the system.
In order for the behavior of the system to be reliable and robust, one wants small fluctuations in the 
output flux while at the same time a small cost of operation. These goals are generally incompatible
as emphasized by a trade-off known under the name of thermodynamic uncertainty relations \cite{Barato2015,Gingrich2016}. 
This trade-off constrains the fluctuations in product formation in enzyme
kinetics \cite{Barato2015d} and can thus be used to infer information on the topology of the 
underlying chemical network \cite{Pietzonka2016,Pietzonka2016a} or to estimate the dissipation
from the fluctuations of observed fluxes 
 \cite{gingrich2017inferring}.
Suppressing fluctuations of an output flux is required to achieve some accuracy with brownian clocks  \cite{Barato2016}, 
while suppressing dissipation leads to an improvement in the thermodynamic efficiency of machines \cite{Pietzonka2016c,Polettini2016,maes2017frenetic,hwang2018energetic,Vroylandt2018}.
This trade-off, originally obtained for non-equilibrium steady states, holds in fact for systems at finite time \cite{pietzonka2017finite,Horowitz2017} evolving in either 
 continuous or discrete time \cite{pigolotti2017generic,proesmans2017discrete,chiuchiu2017mapping}.  
It has also been adapted to Brownian motion \cite{hyeon2017physical}, nonequilibrium self-assembly \cite{Nguyen2016},
active matter \cite{Falasco2016}, equilibrium order parameter fluctuations
\cite{Guioth2016}, phase transitions \cite{solon2017phase}, first-passage-time fluctuations
\cite{garrahan2017simple,Gingrich2017}, and it continues today to generate many new applications or extensions 
\cite{rotskoff2017mapping,dechant2017current,brandner2017thermodynamic,manikandan2018exact,
wierenga2018quantifying,Dechant2018Entropic}.

In this chapter, we focus on the implications of thermodynamic uncertainty relations for chemical kinetics. 
We illustrate the theoretical predictions by studying particle conversion fluxes and their fluctuations 
for both unicyclic and multi-cyclic chemical reactions. We also show the connection with the statistics of first-passage time, defined as the first time that a given number of particles has been converted.

This review is organized as follows. In section 2, we
illustrate bounds on the Fano factor for three examples of 
unicyclic networks, namely the isomerization reaction, the Michaelis-Menten reaction 
and the active catalysis. 
In section 3, we study one example of a multi-cylic network
containing two cycles, which we call the misfolding reaction. 
In section 4, we study complementary relations for the 
 fluctuations of first-passage times. We conclude in section 5.

\section{Bounds for unicyclic networks}

\subsection{The isomerization reaction}\label{sec: isomerization}
Let us consider a single enzyme which can catalyze the transition between 
two isomers $E_1$ and $E_2$:
\begin{center}
\ch{E_1 <>[ $k^{+}$ ][ $k^{-}$  ] E_2},
\end{center}
with constant transition rates $k^+$ and $k^-$.  
This model can be mapped on a biased random walk, with a rate of forward jumps $k^+$ and of backward jumps $k^-$. 
The total displacement of this walker corresponds to 
the difference between the number of isomers $E_1$ converted into $E_2$ minus the number of reverse conversions. 
The master equation of this system is given by 
\begin{equation}\label{master}
    \frac{d}{dt}P_n(t)=k^{-}P_{n+1}(t)-(k^{+}+k^{-})P_n(t)+k^{+}P_{n-1}(t),
\end{equation}
where $P_{n}(t)$ the probability to be at time $t$ at the position $n$.
The detailed balance relation takes the form 
\begin{equation}\label{db}
    \frac{k^{+}}{k^{-}}=e^{\mathcal{A}},
\end{equation}
where $\mathcal{A}$ is the dimensionless affinity given our choice of units, $k_\mathrm{B} T=1$. 

In order to characterize the fluctuations of the variable $n$, we introduce  
the \emph{generating function}: 
\begin{equation}
    \Psi(\lambda,t)=\sum_n e^{\lambda n}P_n(t).
\end{equation}
Using the master equation (\ref{master}), this generation function evolves according to the equation
\begin{equation}\label{gfunction}
    \frac{d \Psi(\lambda,t)}{dt}=\theta(\lambda) \Psi(\lambda,t),
\end{equation}
where we have defined
\begin{equation}\theta(\lambda)\equiv \lim_{t\rightarrow\infty}\frac{1}{t}\ln\left\langle e^{n\lambda}\right\rangle=k^{-}e^{\lambda}-(k^{+}+k^{-})+k^{+}e^{-\lambda}.\end{equation}

The mean and the variance of $n$ can be expressed in terms of the derivatives 
of $\theta(\lambda)$ at $\lambda=0$ as follows:
\begin{eqnarray}
\label{current J}
    J &=& \lim_{t \to \infty} \frac{ \langle n \rangle}{t}=\theta'(0)=k^{+}-k^{-}, \\
    D &=& \lim_{t \to \infty}\frac{\langle n^2 \rangle -\langle n \rangle ^2}{2t}
=\frac{\theta''(0)}{2}=\frac{k^{+}+k^{-}}{2}.
\end{eqnarray}
Using these expressions, Barato \textit{et al}.\cite{Barato2015} have derived an ``uncertainty relation'' involving the following measure of the precision
of the fluctuating variable $n$:
\begin{equation}
    \epsilon ^2=\frac{ \langle n^2 \rangle -\langle n \rangle^2}{\langle n \rangle^2}=
\frac{k^{+}+k^{-}}{(k^{+}-k^{-})^2 t}.
\end{equation}
For a duration $t$, the total energy cost $C$ is the product of 
the entropy production rate by the time $t$, so $C=\mathcal{A} J t$ 
where $J$ is the average conversion rate introduced above.
Then the product of this cost $C$ by the relative uncertainty $\epsilon^2$ can be expressed by means of the detailed balance relation, Eq.~(\ref{db}), as
\begin{equation}\label{uncert}
    C \epsilon^2 =\frac{2D\mathcal{A}}{J}=\mathcal{A}\coth\left(\frac{\mathcal{A}}{2}\right) \geq 2,
\end{equation}
where the last inequality follows from a well known property of hyperbolic tangent.  
Importantly,
 this relation expresses a trade-off between the precision quantified by $\epsilon$ 
and the cost quantified by $C$. 
Note that the inequality in Eq.~(\ref{uncert}) holds arbitrarily far from equilibrium, and becomes  
 saturated only in the linear regime close to equilibrium when $\mathcal{A} \to 0$. 

\subsection{The reversible Michaelis-Menten reaction} \label{sec:MM}
We now consider another important unicyclic network, namely the well-known Michaelis-Menten kinetics \cite{Barato2015d}. 
In this chemical network, a substrate $S$ is transformed into a product $P$ 
due to the presence of an enzyme $E$ via the formation of an unstable complex $ES$: 
\begin{center}
\ch{S + E <>[ $k_{1}^{+}$ ][ $k_{1}^{-}$  ] ES <>[ $k_{2}^{+}$  ][ $k_{2}^{-}$ ] P + E},
\end{center}
where $k_{1}^{+}$ is proportional to the substrate concentration and $k_{2}^{-}$ is proportional to the 
product concentration. The local detailed balance relation is now given by
\begin{equation}\label{db1}
    \frac{k_{1}^{+}k_{2}^{+}}{k_{1}^{-}k_{2}^{-}}=e^\mathcal{A}.
\end{equation}
We introduce $p_{\alpha,n}(t)$ as the probability to have the enzyme in the state $\alpha=0,1$ with 
$\alpha=0$ representing the free state and $\alpha=1$ the bound state, and with $n$ molecules of P produced. 
This probability satisfies the master equations:
\begin{eqnarray}
\frac{dp_{ 0,n}}{dt}= k_{1}^{-}p_{1,n} + k_{2}^{+}p_{1,n-1} -(k_{1}^{+} + k_{2}^{-})p_{ 0,n}; \\
\frac{dp_{ 1,n}}{dt}= k_{1}^{+}p_{0,n} + k_{2}^{-}p_{1,n+1} -(k_{1}^{-} + k_{2}^{+})p_{ 1,n}.\label{MEMM}
\end{eqnarray}
We introduce again generating functions associated to these probability distributions by 
\begin{equation}
    \Psi_{\alpha}(\lambda, t)=\sum _{n=-\infty}^{+\infty} e^{\lambda (n+\alpha/2)} p_{\alpha,n}(t).
\end{equation}
By convention a half integer value of $n$ is assigned to states where the enzyme is bound, 
and a integer number when the enzyme is free. By transforming the master equation into an 
evolution equation for the generating function we find:
\begin{equation}
\frac{d}{dt} \begin{pmatrix}\Psi_0\\\Psi_1 \end{pmatrix} = L(z)\begin{pmatrix}\Psi_0\\\Psi_1 \end{pmatrix},
\end{equation}
with the evolution matrix 
\begin{equation}
L(z)=\begin{pmatrix}
    -(k_{1}^{+}+k_{2}^{-})       & z^{-1}k_{1}^{-}+zk_{2}^{+} \\
    zk_{1}^{+}+z^{-1}k_{2}^{-}      &  -(k_{1}^{-}+k_{2}^{+})  \end{pmatrix},
\end{equation}
where $z=e^{\lambda/2}$.
By the Perron-Frobenius theorem, there is a non-degenerate positive leading eigenvalue of $L$, 
which we denote 
$\theta(z(\lambda))$. 
 Explicit evaluation of this function yields 
\begin{eqnarray}\label{current}
 J &=& \left. \frac{d \theta}{d \lambda} \right|_{\lambda=0}=\frac{k_{1}^{+}k_{2}^{+}-k_{1}^{-}k_{2}^{-}}{k_{1}^{+}+k_{1}^{-}+k_{2}^{+}+k_{2}^{-}};\\ 
D &=& \frac{1}{2} \left. \frac{d^2 \theta}{d^2 \lambda} \right|_{\lambda=0} =\frac{k_{1}^{+}k_{2}^{+}+k_{1}^{-}k_{2}^{-}-2J^2}{2(k_{1}^{+}+k_{1}^{-}+k_{2}^{+}+k_{2}^{-})}.
\end{eqnarray}

In the context of enzymatic kinetics, the \emph{Fano factor} characterizes 
the fluctuations in the formation of product by the enzyme. It is defined as
\begin{equation}
F=\frac{2D}{J}.
\end{equation}
Using the above expressions for $J$ and $D$ together with the detailed balance condition (\ref{db1}) we find that
\begin{equation}\label{fanom}
F=\coth\left(\frac{\mathcal{A}}{2}\right)-\frac{2k_{2}^{-}k_{1}^{-}(e^{\mathcal{A}}-1)}{(k_{2}^{-}+k_{2}^{+}+k_{1}^{-}+k_{1}^{+})^2}.
\end{equation}
A lower bound for the Fano factor can be obtained by minimizing the right-hand side of Eq.~(\ref{fanom}) 
with respect to all the transition rates. The minimum is obtained when
$k_{1}^{+}=k_{2}^{+}\equiv k^{+}$, and $k_{1}^{-}=k_{2}^{-}\equiv k^{-}$.
Then, using the detailed balance condition, one proves  
the inequality \cite{Barato2015d}
\begin{equation}
\label{MM}
F\geq \frac{1}{2}\coth\left(\frac{\mathcal{A}}{4}\right)\geq\frac{2}{\mathcal{A}}. 
\end{equation}

\begin{figure}[!htb]
\begin{center}
\includegraphics[scale=0.25]{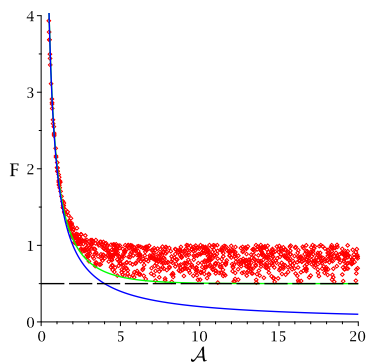}
\end{center}
\caption{The Fano factor $F$ as a function of the affinity $\mathcal{A}$ in the Michaelis-Menten kinetics. 
The black horizontal dashed line represents $F_{min}=0.5$, the blue line the bound $2/\mathcal{A}$, 
the green line is the hyperbolic bound of Eq.~(\ref{MM}).}
\label{fig:L1}
\end{figure}

Since the relation $C\epsilon^2=F\mathcal{A}$ holds generally, both 
Eq.~(\ref{uncert}) and Eq.~(\ref{MM}) are particular cases of a general  
inequality for $C \epsilon^2$ valid for a unicyclic enzyme containing $N$ states. 
In their original work \cite{Barato2015}, Barato \textit{et al.} provided an inequality involving, 
$\mathcal{A}$ and $N/n_c$ where $n_c$ was 
defined as the number of consumed substrate molecules in each cycle.
If one defines the Fano factor per molecule instead of per cycle, as we do here, there is no need for $n_c$, and the result is simpler to state: 
the Fano factor of a unicyclic enzyme containing $N$ states is bounded by  
an expression that only depends on $\mathcal{A}$ and $N$ \cite{Barato2016}:
\begin{equation}\label{bound1}
    F \geq \frac{1}{N} \coth\left(\frac{\mathcal{A}}{2N}\right).
\end{equation}
The example we gave previously of an isomerization reaction satisfies this relation with $N=1$,
while the Michaelis-Menten scheme does so with $N=2$. 
In Fig.~\ref{fig:L1}, we verify this bound by computing the Fano factor for
1000 random transition rates of the form $k_1^+=10^{2 r_1-1}$ 
where $r_1$ is a uniform random number in $[0,1]$, while keeping a fixed value of the affinity $
\mathcal{A}
$. 
 
An affinity-independent bound follows from Eq.~(\ref{bound1})
by letting $\mathcal{A} \to \infty$ \cite{moffitt2010mechanistic,knoops2017motion}
\begin{equation}
    F\geq \frac{1}{N}.
\label{1/N}
\end{equation}
The limit $\mathcal{A} \to \infty$ is realized in practice as soon as one transition contributing to the 
affinity $\mathcal{A}$ becomes irreversible.
Note that Eq.~(\ref{1/N}) can also be derived from an analysis of current fluctuations in a periodic 
one dimensional lattice \cite{Derrida1983}. It represents a central result of statistical kinetics, since it allows to 
estimate the number of states in an enzymatic cycle from measurements of fluctuations \cite{Moffitt2014}.
  
As shown in Fig.~\ref{fig:L1}, all the points are indeed above 
the bound $F_{\min}=0.5$ in the case of Michaelis-Menten kinetics. 
In order to explore further Fano factor bounds, we now move to more complex examples.

\subsection{The active catalysis} 
\label{sec:active catalyst}
\begin{figure}[!h]
\begin{center}
\includegraphics[scale=0.45]{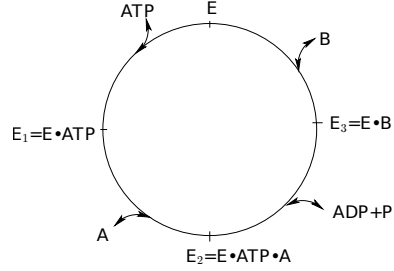}
\end{center}
\caption{Cycle representation of the reactions involved in the active catalysis}
\label{fig:active-cat}
\end{figure}

In this type of chemical reaction, the folding of the substrate molecule $A$ 
into the product molecule $B$ 
is accompanied by the hydrolysis of an ATP molecule. 
This reaction can be represented as a unicyclic network with four intermediate states $E$, $E_1$, 
$E_2$ and $E_3$ for the enzyme and two substrates $A$ and $B$ as 
shown in Fig.~\ref{fig:active-cat}.
The various reactions are  
\begin{center}
\ch{ATP + E <>[ $k_{1}^{+}$ ][ $k_{1}^{-}$  ] E_1, \, \, E_1 + A <>[ $k_{2}^{+}$  ][ $k_{2}^{-}$ ] E_2},
\end{center}
\begin{center}
\ch{E_2 <>[ $k_{3}^{+}$  ][ $k_{3}^{-}$ ] E_3 + ADP + P, \, \, E_3 <>[ $k_{4}^{+}$  ][ $k_{4}^{-}$ ] E + B.}
\end{center}

The local detailed balance condition 
takes the form 
\be
\frac{k_1^+ k_2^+ k_3^+ k_4^+}{k_1^- k_2^- k_3^- k_4^-}=e^\mathcal{A},
\ee
where the affinity of the cycle now reads $\mathcal{A}=\mu_A - \mu_B + \Delta \mu$,
in terms of $\mu_A$ (resp.~$\mu_B$) the chemical potential of the substrate $A$ (resp.~$B$) and 
$\Delta \mu$ the chemical potential difference associated with the ATP hydrolysis reaction.
 

The framework of the previous section applies again here: 
now the evolution of the generation function is governed by a $4\times4$ matrix, and for this reason there is no simple analytic expression for its eigenvalues. Nevertheless, it is still possible to compute the corresponding currents and diffusion coefficients without having to obtain these explicitly by exploiting a method due to Koza as explained in Ref.~ \cite{koza2002maximal,Barato2015d}. 

\begin{figure}[!htbp]
\begin{center}
\includegraphics[scale=0.25]{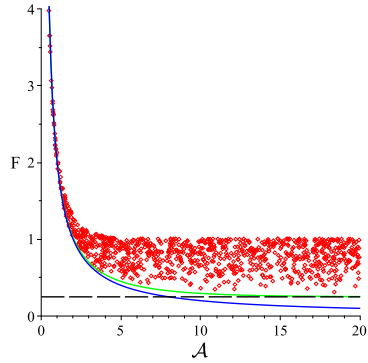}
\end{center}
\caption{The Fano factor $F$ as a function of the affinity $\mathcal{A}$ for the active catalyst kinetics. 
The black horizontal dashed line represents $F_{min}=0.25$, the blue line the bound $2/\mathcal{A}$, 
the green line is the hyperbolic bound of Eq.~(\ref{bound1}) with $N=4$.}
\label{fig:L2}
\end{figure}

By following this method, we obtain the plot shown in Fig.~\ref{fig:L2}.
As shown in this figure, the bound satisfies to the general property expected 
for a unicyclic enzyme given in Eq.~(\ref{bound1}) with $N=4$ intermediate states.

\section{Bounds for multi-cyclic networks: the misfolding reaction}\label{sec:misfolding}
We now switch to multi-cyclic reaction networks. 
As a simple example we consider the misfolding reaction, which describes an enzyme that can make errors. More precisely, the enzyme can bind a molecule $A$ and lead to the production of the ``correct" molecule, say, $B$, or a ``wrong" one, say $C$. This scheme represents a network with two cycles, characterized by the same free and bound state of the enzyme. The two possible reactions that can occur are:
\begin{center}
\ch{A + E <>[ $k_{1}^{+}$ ][ $k_{1}^{-}$  ] E^* <>[ $k_{2}^{+}$  ][ $k_{2}^{-}$ ] B + E},
\end{center}
\begin{center}
\ch{A + E <>[ $k_{1}^{+}$ ][ $k_{1}^{-}$  ] E^* <>[ $k_{3}^{+}$  ][ $k_{3}^{-}$ ] C + E}.
\end{center}

We now have two different affinities driving each one of these reactions, defined by:
\begin{equation}\label{db2}
\begin{split}
 \frac{k_{1}^{+}k_{2}^{+}}{k_{1}^{-}k_{2}^{-}}&=e^{\mathcal{A}_1}; \\
  \frac{k_{1}^{+}k_{3}^{+}}{k_{1}^{-}k_{3}^{-}}&=e^{\mathcal{A}_2}.
  \end{split}
\end{equation}
The master equations describe the evolution of the probability of being in the two different enzyme states (bound and free) 
as a function two integer chemical variables $n$ (resp.~$m$), which represent the number 
of $B$ (resp.~$C$) produced since an arbitrary time. 
We then have
\begin{equation}
\begin{split}
    \frac{dp_0(n,m,t)}{dt}&=k_1^{-}p_1(n, m, t)+k_{2}^{+}p_1(n-1,m,t)+k_{3}^{+}p_1(n,m-1,t) \\ &\qquad {}-(k_{1}^{+} +k_{2}^{-} +k_{3}^{-})p_0(n, m, t) ; \\
    \frac{dp_1(n,m,t)}{dt}&=k_1^{+}p_0(n, m, t)+k_{2}^{-}p_0(n+1, m,t)+k_{3}^{-}p_0(n,m+1,t)\\&\qquad {}-(k_{1}^{-} +k_{2}^{+} +k_{3}^{+})p_1(n, m, t).
 \end{split}
\end{equation}
The generating function for this system can be defined by 
\begin{equation}
    \Psi_{\alpha}(\overline{\lambda},t)=\sum_{n,m}e^{\overline{\lambda} \cdot (n+\alpha/2,m+\alpha/2)}p_{\alpha}(n,m,t),
\end{equation}
where $\overline{\lambda}$ is a vector containing the two variables $\lambda_1$ and $\lambda_2$ associated to the degrees of freedom $n$ and $m$ respectively. The evolution matrix that governs the dynamics of the generating function is given by
\begin{equation}
    L(z_1,z_2)=\begin{pmatrix}
    -(k_{1}^{+}+k_{2}^{-}+k_{3}^{-})       & (z_1z_2)^{-1}k_{1}^{-}+\frac{z_1}{z_2}k_{2}^{+}+\frac{z_2}{z_1}k_{3}^{+} \\
    z_1z_2k_{1}^{+}+\frac{z_2}{z_1}k_{2}^{-}+\frac{z_1}{z_2}k_{3}^{-}      &  -(k_{1}^{-}+k_{2}^{+}+k_{3}^{+})  \end{pmatrix}
\end{equation}
Again, the leading eigenvalue of $L(z_1,z_2)$, called $\Theta(z_1,z_2)$, allows us to obtain 
the currents $J_1$ and $J_2$ and their diffusion coefficients $D_1$ and $D_2$. 

Using these parameters, we can compute two Fano factors $F_1$ and $F_2$, 
defined by $F_i=2D_i/J_i$, 
where $i=1,2$. Similarly, we define the corresponding cost-fluctuations parameters
$C \epsilon_i^2=F_i \Sigma/J_i$, which instead contain the total entropy production rate
$\Sigma=\mathcal{A}_1 J_1+\mathcal{A}_2 J_2$.


The error-cost parameters $C \epsilon_i^2$ 
are constrained by the uncertainty relation \citep{Barato2015d} :
\be
C \epsilon_i^2 \ge \frac{\bar{\mathcal{A}}}{2} \coth\left( \frac{\bar{\mathcal{A}}}{4} \right) 
\ge \max(2,\frac{\bar{\mathcal{A}}}{2}),
\label{Ceps-multicyclic}
\ee
where $\bar{\mathcal{A}}$ is the minimum of the two cycle affinities $\mathcal{A}_1$ and $\mathcal{A}_2$.

\begin{figure}[!htb]
\begin{center}
\includegraphics[scale=0.25]{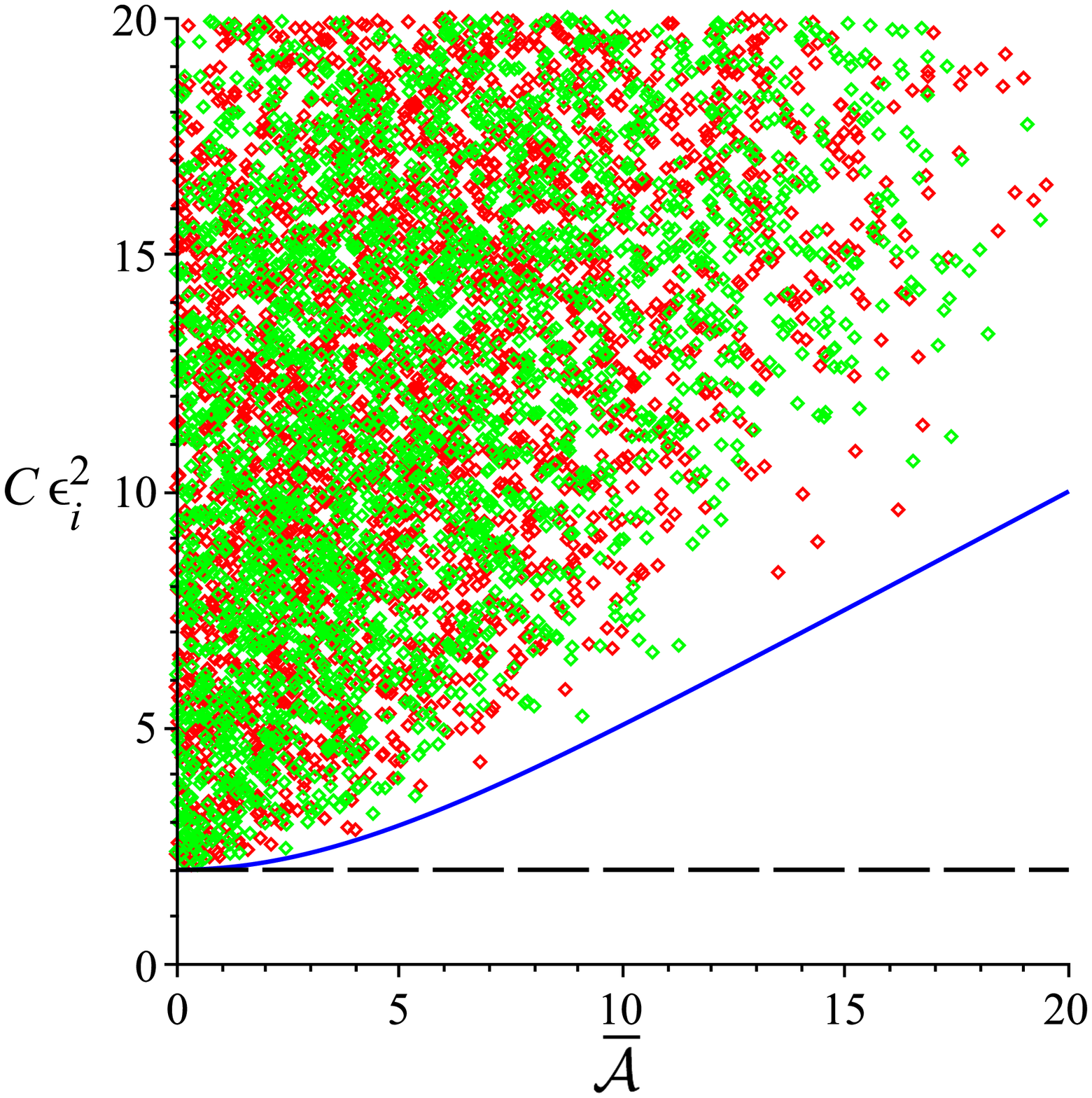}
\includegraphics[scale=0.25]{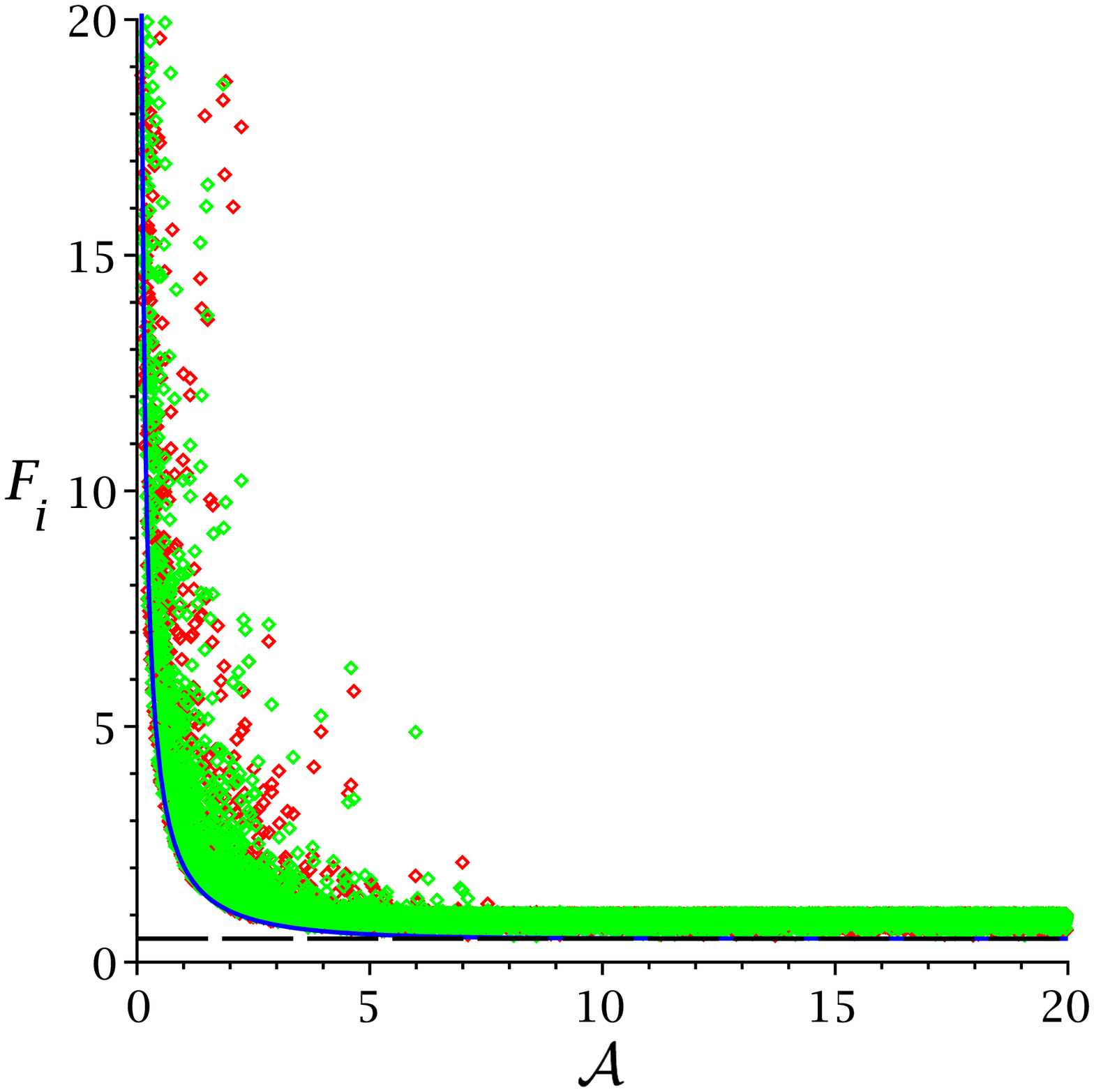}
\end{center}
\caption{Left: Error-cost parameters $C \epsilon_i^2$ as a function of the 
minimum $\bar{\mathcal{A}}$ of the two different affinities $\mathcal{A}_1$ and $\mathcal{A}_2$ of the two cycles. The 
blue solid and black dashed line illustrate the bound of Eq.~(\ref{Ceps-multicyclic}).
Right: Fano factor $F_i$ as function of the single affinity $\mathcal{A}=\mathcal{A}_1=\mathcal{A}_2$. 
The blue solid and black dashed line illustrate the bound of Eq.~(\ref{bound1}). 
In both figures, red points 
represent $i=1$ and green points $i=2$.}
\label{fig:L3}
\end{figure}

We have verified numerically this inequality in the left panel of Fig.~\ref{fig:L3}, 
which was constructed by generating randomly transition rates for 5000 iterations and then 
evaluating the error-cost parameters and the two different cycle affinities $\mathcal{A}_1$ and $\mathcal{A}_2$ 
using the detailed balance conditions (\ref{db2}). 

In this case where the affinities $\mathcal{A}_1$ and $\mathcal{A}_2$ differ, we observe
that a bound of the form of Eq.~(\ref{bound1}) does not hold in general 
for the two Fano factors $F_i$ in terms of the affinities $\mathcal{A}_i$ or $\bar{\mathcal{A}}$.
In contrast to that, the two Fano factors $F_i$ are always bounded by $1/2$, 
which is the limit expected for unicycles with $N=2$ states. 
This confirms that the bound in $1/N$ for the Fano factor---a central result in statistical 
kinetics---holds in general and is not just a consequence of the thermodynamic uncertainty relation.

In the particular case where the two affinities $\mathcal{A}_1$ and $\mathcal{A}_2$ are equal, 
then the bound of the form of Eq.~(\ref{bound1}) holds again for the two Fano factors $F_i$ 
as shown in the right panel of Fig.~\ref{fig:L3}. Thus, we have illustrated the bound 
for the error-cost parameter which is linked to the the uncertainty relation, but 
importantly we find that no affinity dependent 
bound of this type exist for the two Fano factors except in
 the particular case where all the cycles have the same affinity.

\section{Bound on the fluctuations of first-passage times}
The Fano factor bounds derived above represent a general property 
of current fluctuations probed for a fixed observation time. It is also interesting to 
look at these results from a different point of view, which 
focuses instead on fluctuations of first-passage times to reach 
a threshold of time-integrated current \cite{Gingrich2017}.

In this section, we derive the corresponding relations for  
the examples we have studied above.
We rely on the so-called renewal equation to 
analyze first-passage times for Markovian jump processes. This equation connects the propagator $p(n, t)$, \textit{i.e.},
the probability of being in $n$ at time $t$ given that one starts from 0 at time 0,
to the probability $F(n,t)$ to reach the state $n$ for the first time 
at time $t$ \cite[p.~307]{VanKampen2007_vol}:
\begin{equation}
    p(n,t)=\delta_{n,0}\delta_{t,0}+\int_0^td\tau\,F(n,\tau)p(0,t-\tau),
\end{equation}
where one assumes that the initial state is at $n=0$ and that the process is translation invariant.
Taking the Laplace transform of this equation gives
\begin{equation}
\label{renewal}
    \tilde{F}(n,s)=\frac{\tilde{p}(n,s)}{\tilde{p}(0,s)},
\end{equation}
with $\tilde{F}(n,s)$ (resp.~$\tilde{p}(x,s)$) the Laplace transforms of 
$F(n,t)$ (resp.~$p(n,t)$). Since $\tilde{F}(x,s)$ is the moment generating function of the first-passage time to reach $n$, 
which we now denote $T$, the first and second cumulants of $T$ are:
\begin{equation}
\label{cumul12}
    \left\langle T\right\rangle=-\left.\frac{d}{ds}\ln\left(\tilde{F}(n,s)\right)\right|_{s=0},\qquad \textrm{Var}(T)=\left.\frac{d^2}{ds^2}\ln\left(\tilde{F}(n,s)\right)\right|_{s=0}.
\end{equation}

The connection between current fluctuations probed for a fixed observation time and 
first-passage times fluctuations goes in fact beyond the first and second moments.
Indeed, let us evaluate the cumulant generating function of the first-passage time $T$:
\be
    g(s)= \lim_{n \to \infty} \frac{1}{n}\ln\left\langle e^{-s T}\right\rangle = 
     \lim_{n \to \infty} \frac{1}{n} \ln \tilde{F}(n,s)\label{cgffpt}
\ee
One can then show that $g(s)$ is related to the cumulant generating function 
of the flux of $n$, which is the function we have denoted earlier by $\theta(s)$, by the following relation\cite{Gingrich2017}:
\begin{equation}
    g(s)=\theta^{-1}(s).
    \label{inversion}
\end{equation}

\subsection{Isomerization relation}
Let us now verify these new relations, starting with the isomerization reaction.  
In this case, the equation obtained by Laplace transforming the master equation
 Eq.~(\ref{master}) admits a solution of the form $\tilde{p}_n(s)=\mathcal{N} \lambda(s)^n$ when $W^+ \ge W^-$ and $n \ge 0$, with
\begin{equation}
\lambda(s)=\frac{s+W^+ + W^- - \sqrt{(s+W^+ + W^- )^2-4 W^+ W^-}}{2 W^+}.\label{lam}
\end{equation} 
Using Eq.~(\ref{renewal}), one obtains $\tilde{F}(n,s)=\lambda(s)^n$. Then with Eq.~(\ref{cumul12}), one finds
\begin{equation}\left\langle T\right\rangle=\frac{n}{W^+-W^-},\qquad \textrm{Var}(T)=\frac{n (W^++W^-)}{(W^+-W^-)^3}.\end{equation}
We recall that the energy cost $C$ is related to the entropy production rate $\Sigma$ by $C=t \Sigma=\mathcal{A} J t$, with 
the affinity $\mathcal{A}$ defined in Eq.~(\ref{db}) and the average current $J$ defined in Eq.~(\ref{current J}). 
Then, one obtains the uncertainty relation complementary to Eq.~(\ref{uncert}):
\begin{equation}
\Sigma \, \frac{\textrm{Var}(T)}{\left\langle T\right\rangle}=\mathcal{A}\coth\left(\frac{\mathcal{A}}{2}\right) \geq 2.\label{turfp}
\end{equation} 
This relation means that fluctuations in first-passage times can be reduced only 
at the price of an increase of dissipation. 
Note that fluctuations of first-passage times are not constrained by dissipation  
when $W^+ \le W^-$, because in this case the mean first-passage time is infinite.  
 
The relation between the generating functions of first-passage time and current is also easily verified. 
Indeed, since $\tilde{F}(n,s)=\lambda(s)^n$, 
\be
g(s)=  \ln \lambda(s).
\label{g-fct}
\ee
Then, from the definition of the cumulant generating function of $n$ introduced 
in Eq.~(\ref{gfunction}), one finds
\begin{equation}
    s=\theta[ \theta^{-1}(s) ]= W^+ e^{\theta^{-1}(s)}+W^- e^{-\theta^{-1}(s)}-(W^+ + W^-),
\end{equation}
therefore
\begin{equation}
     \theta^{-1}(s)=\ln \left( \frac{s+W^+ + W^- -\sqrt{(s+W^- + W^+)^2-4 W^- W^+}}{2W^+} \right),
\end{equation}
which is clearly equivalent to plugging Eq.~(\ref{lam}) into Eq.~(\ref{g-fct}) in agreement with Eq.~(\ref{inversion}). 

\subsection{Michaelis-Menten reaction} 
The first-passage time uncertainty relation can be validated in an analogous way for the Michaelis-Menten reaction. We first need to determine $\tilde{p}_{0/1,n}(n,s)$. This can be done by taking the Laplace transform of Eq.~(\ref{MEMM}):
\begin{eqnarray}
    (s+k^+_1+k^-_2)\tilde{p}_{0,n}&=k^-_1 \tilde{p}_{1,n}+k^+_2\tilde{p}_{1,n-1},\\
(s+k^-_1+k^+_2)\tilde{p}_{1,n}&=k^+_1 \tilde{p}_{0,n}+k^-_2\tilde{p}_{0,n+1}.
\end{eqnarray}
Along the lines of the isomerization reaction, we assume that $\tilde{p}_{0/1,n}(s)=\mathcal{N}_{0/1}\lambda(s)^n$, leading to
\begin{eqnarray}
\lambda(s)=\frac{s^2+Ks+k^++k^--\sqrt{\left(s^2+Ks+k^++k^-\right)^2-4k^+k^-}}{2k^-},\label{lamMM}
\end{eqnarray}
where we have introduced
\begin{equation}
K=k_1^++k_2^++k_1^-+k_2^-,\qquad k^-=k_1^-k_2^-,\qquad k^+=k_1^+k_2^+
\end{equation}
to simplify notations. This again leads to the first two cumulants:
\begin{eqnarray}
\left\langle T\right\rangle&=&\frac{Kn}{k^+-k^-},\qquad
\textrm{Var}(T) =\frac{\left(K^2(k^++k^-)-2(k^+-k^-)^2\right)n}{(k^+-k^-)^3}.\label{VarMM}
\end{eqnarray}
The thermodynamic uncertainty relation for first-passage times, Eq.~(\ref{turfp}) can now be verified easily.

As mentioned before, the cumulant generating function associated with the first-passage time is given by $g(s)=\ln \lambda(s)$, where $\lambda(s)$ is given by Eq.~(\ref{lamMM}).
One can invert this expression to determine the cumulant generating function $\theta(\mu)$ associated with the number of produced particles:
\begin{equation}
    \theta(s)=g^{-1}(s)=\frac{\sqrt{K^2+4(e^{s}-1)(k^--k^+e^{-s})}-K}{2}
\end{equation}

\subsection{Misfolding reaction} 
As a final example, we shall now turn to the misfolding reaction. 
This reaction network can be decomposed into two independent fluxes: the production of $B$ molecules and the production 
of $C$ molecules. Let us focus on the first-passage time to produce $n$ molecules of $B$ type. 
This problem can be mapped 
on the Michaelis-Menten reaction: indeed $B$ is produced from $E^*$ at a rate $k_2^+$ and produces 
$E^*$ at a rate $k_2^-$. On the other hand, $E^*$ is constructed from some other source (either $A$ or $C$) 
at the rate ${k'}_1^+=k_1^++k_3^-$ and deconstructed at rate ${k'}_1^-=k_1^-+k_3^+$. Therefore the system can be
 mapped onto a Michaelis-Menten system with $k_1$ replaced by ${k'}_1$. One concludes that
 Eqs.~(\ref{lamMM}--\ref{VarMM}) also hold for the misfolding reaction, with the appropriate 
change of rates. Using the expression for the entropy production rate $\Sigma$ determined in Section~\ref{sec:misfolding}, leads to the thermodynamic uncertainty relation in the form:
\begin{equation}
    \Sigma \frac{\textrm{Var}(T)}{\left\langle T\right\rangle}\geq 2.
\end{equation}

\section{Conclusion}

In this chapter, we have illustrated
a number of thermodynamic bounds 
for chemical kinetics and particularly for chemical cycles.
In both unicyclic and multicyclic networks, we have confirmed 
the thermodynamic uncertainty relation which limits the precision that a chemical 
system can achieve for a given cost in terms of chemical dissipation.
We have pointed out that only in unicyclic networks or in multicyclic networks
subjected to a single affinity, there is a simple 
 affinity dependent bound. In contrast to that, there is always  
an affinity independent bound for the Fano in terms of 
the inverse number of states, but this bound does not contain
any trade-off.

Very recently, Gingrich and Horowitz reported a relation
between the large deviation functions for currents
and first-passage times in general Markov chains \cite{Gingrich2017}. They also made an
interesting connection between the thermodynamic uncertainty relation
and first-passage time statistics.
In this chapter, we have also verified their result on our examples. 
In future work, we would like to explore this connection further, because
it could be used in both ways: on one 
hand one could gain insights into currents fluctuations using results on 
first-passage time statistics and on the other hand one can understand better 
first-passage time statistics using large-deviation techniques, 
originally introduced for the analysis of current fluctuations
in non-equilibrium systems.

\section*{Acknowledgments}
LP acknowledges support from a Chair of the Labex CelTisPhysBio (Grant No. ANR-10-LBX-0038). KP was supported by the Flemish Science Foundation
(FWO-Vlaanderen) travel grant V436217N.

\bibliographystyle{ws-rv-van.bst}

\end{document}